\documentclass[12pt,preprint]{aastex}

\begin{document}

\title{Spectral Softening in X-ray Afterglow of GRB~130925A as Predicted by Dust Scattering Model}

\author{\sc Yi-Nan Zhao\altaffilmark{1}, and Lang Shao\altaffilmark{1,2\star}}
\altaffiltext{1}{Department of Space Science and Astronomy, Hebei Normal University, Shijiazhuang 050024, China;}
\altaffiltext{2}{Key Laboratory of Dark Matter and Space Astronomy, Purple Mountain Observatory, Chinese Academy of Sciences, Nanjing 210008, China}
\altaffiltext{$\star$}{lshao@hebtu.edu.cn}

\begin{abstract}
Gamma-ray bursts (GRBs) usually occurs in a dense star-forming region with massive circum-burst medium. The small-angle scattering of intense prompt X-ray emission off the surrounding dust grains will have observable consequences, and sometimes can dominate the X-ray afterglow. In most of the previous studies, only Rayleigh-Gans (RG) approximation is employed for describing the scattering process, which works accurately for the typical size of grains (with radius $a\leq 0.1\,{\rm \mu m}$) in the diffuse interstellar medium. When the size of the grains may significantly increase as in a more dense region where GRBs would occur, the RG approximation may not be valid enough for modeling detailed observational data. In order to study the temporal and spectral properties of the scattered X-ray emission more accurately with potentially larger dust grains, we provide a practical approach using the series expansions of anomalous diffraction (AD) approximation based on the complicated Mie theory. We apply our calculations to understanding the puzzling X-ray afterglow of recently observed GRB~130925A which showed a significant spectral softening. We find that the X-ray scattering scenarios with either AD or RG approximation adopted could both well reproduce the temporal and spectral profile simultaneously. Given the plateau present in early X-ray light curve, a typical distribution of smaller grains as in the interstellar medium would be suggested for GRB~130925A.
\end{abstract}
\keywords{dust, extinction --- gamma rays: bursts --- interstellar
medium --- scattering --- X-rays: general}

\section{Introduction}
\label{sec:intro}

The nature of interstellar dust grain has been well studied (Draine 2003 and the reference therein), while the dust grains around GRBs are still poorly known. Nevertheless, the existence of dust grains around GRBs has been well established. The commonly found optically ``dark bursts'' have been mainly diagnosed as the result of dust extinction in the host galaxy (Lazzati et al., 2002; Perley et al. 2009; Greiner et al. 2011). The spectral energy distributions of optical/near-infrared afterglows that deviate from the intrinsic power law are also indicative of dust extinction in the host galaxy (Stratta et al. 2004; Kann et al. 2006; Chen et al. 2006). Another important indication is the intrinsic excess of gas column density in X-ray spectra (Stratta et al. 2004; Campana et al. 2006; Campana et al. 2012).

X-ray scattering by dust grains is an important tool for investigating the physical properties of the interstellar medium along the line of sight (Mathis \& Lee 1991; Predehl \& Klose 1996; Smith \& Dwek 1997; Draine \& Tan 2003). Since gamma-ray bursts (GRBs) are usually in star-forming regions which are rich in dust grains, several pioneering studies have investigated the idea that the X-ray flux from a GRB could be affected by the circum-burst dust grains (Klose 1998; M\'{e}sz\'{a}ros \& Gruzinov 2000; Sazonov \& Sunyaev 2003). Shao \& Dai (2007) provided the first treatment for evaluating both the temporal and spectral evolution of the delayed emission due to X-ray scattering by circum-burst dust grains in an illustrative model. Interestingly, the quite puzzling light curves, in particular, a shallow decay followed by a ``normal'' decay and a further rapid decay of X-ray afterglows can be well understood with such a simple model (Shao et al. 2008). This model, so called as the dust scattering model, predicted a strong spectral softening in the X-ray afterglow and was not supported by the observational data of the time back then (Shen et al. 2009). Even though most X-ray afterglows did not exhibit such a spectral softening, it would be a great challenge to the standard afterglow models if this feature indeed showed up in some X-ray afterglows as in GRB~090417B (Holland et al. 2010). We will investigate this spectral feature further in this paper.

The scattering of X-ray photons by the interstellar dust was first studied using the Rayleigh-Gans (hereafter RG) approximation for spherical grains (Overbeck 1965). The validity of the RG approximation requires a limiting case of the phase shift across the grain, i.e., $|\rho|=2x|m-1| \approx 6\,(E/1\,{\rm keV})^{-1}(a/ 1\,{\rm \mu m})\ll1$ (see the Appendix B for details), where $x=2\pi a/\lambda$, $a$ is the radius of the dust grain, $\lambda$ is the wavelength of the photon, $E$ is the energy of the photon, and $m$ is the complex refractive index of the dust grain (van de Hulst 1957; Alcock and Hatchett 1978).
Even though commonly employed, the RG approximation is no longer valid at X-ray energies $E\lesssim 1 {\rm KeV}$ for typical interstellar dust grains with an average radius $\bar{a}\sim 0.1\,{\rm \mu m}$ (Smith and Dwek 1998). This was also cautioned in Shao \& Dai (2007) where the typical interstellar dust model (e.g., Mathis et al. 1977) was considered when using RG approximation for X-ray Scattering around GRBs. Figure 1 shows the valid regions of different photon energy and grain size for RG approximation and anomalous diffraction (hereafter AD) approximation, respectively.

The size distribution of interstellar dust grain in the Milky Way Galaxy has been well determined by reproducing the Galactic extinction curves (Weingartner and Draine 2001), while it has not been well constrained around GRBs. Even though the GRB extinction curves have been found complicated as a whole (Zafar et al. 2011; Kr\"{u}hler et al. 2011; Schady et al. 2012; Covino et al. 2013), a more realistic size distribution favoring larger dust grains has been suggested by flatter extinction curves derived from optical/near-infrared spectra of some GRBs (Stratta et al. 2004; Chen et al. 2006; Li et al. 2008; Liang \& Li 2009; 2010). This trend might be consistent with the fact that most GRBs are probably in a dense star-forming region where small dust grains tend to coagulate onto large ones (Jura 1980; Kim 1996; Maiolino et al. 2001; Weingartner \& Draine 2001), or the fact that the radiation field around GRBs is strong so that smaller grains tend to be destroyed more rapidly (e.g., Waxman \& Draine 2000). The extinction to observed ratio of hydrogen column density, $A_V/N_{\rm H}$, is one measure to distinguish these mechanisms. The coagulation model predicts a reduction in $A_V/N_{\rm H}$ and the accretion of gas-phase material can predict an enhancement in $A_V/N_{\rm H}$ (Whittet 2003).

In this paper, to better understand the spectrally softening feature predicted by the dust scattering model, we investigate the X-ray scattering scenario with the AD approximation which relaxes the condition of ``phase shift'' previously violated by using the RG approximation. In particular, we compare the temporal and spectral features obtained with these two approximations and apply our results to understanding the X-ray afterglow of GRB~130925A that also showed the spectral softening that is consistent with the model prediction. The paper is structured as follows. In Section 2, we describe a practical approach for evaluating the differential scattering cross section using the series expansions of the AD approximation. In Section 3, we investigate the temporal and spectral properties of dust scattering model employing AD approximation, compared with those of RG approximation. In Section 4, we apply the model to understanding the X-ray afterglow of GRB~130925A. In Section 5, we have a summary of our conclusions.

\section{Anomalous Diffraction Approximation}
\label{sec:ad}

For X-ray scattering off typical astrophysical dust grains, we have $x=2\pi a/\lambda\gg1$ and $|m-1|\ll1$, so the complex refractive index $m$ and phase shift $\rho$ can be approximated as real values (Draine \& Lee 1984; Laor \& Draine 1993; L\u{u} et al. 2011). Therefore, the differential cross section
can be given by
\begin{eqnarray}
\frac{{\rm d}\sigma_{{\rm sca}}}{{\rm d}\Omega}=k^2 a^4 |A|^2,
\label{eq:dcs}
\end{eqnarray}
where $k$ is the wave number of the photon, and the complex amplitude function $A$ which, in the AD approximation (also referred to as the AD theory in the literature, e.g., Draine \& Tan 2003; Draine \& Allaf-Akbari 2006), can be give in the integral form by (van de Hulst 1957)
\begin{eqnarray}
A=\int_{0}^{\frac{\pi}{2}}\left(1-{\rm e}^{-i\rho\sin\tau}\right)J_{0}\left(z\cos\tau\right)
\cos\tau \sin\tau{\rm d}\tau,
\label{eq:int}
\end{eqnarray}
where $z\equiv2\pi a\theta/\lambda$, $\theta$ is the scattering angle, and $J_{\alpha}(x)$ is the Bessel function of the first kind.
Here we can simplify this equation, the imaginary part can be given by
\begin{eqnarray}
{\rm Im}A=\frac{\rho}{y^2}\left(\frac{\pi y}{2}\right)^{\frac{1}{2}}
J_{\frac{3}{2}}\left(y\right),
\label{eq:im}
\end{eqnarray}
where we define $y\equiv\sqrt{ \rho^2+z^2 }$.
According to van de Hulst (1957), for small $\rho$, the approximate series expansion of the real part of the amplitude function is
\begin{eqnarray}
{\rm Re}A_{{\rm small}}=\sum_{n=1}^{\infty}\frac{{(-1)}^{n+1}2^n n!}{(2n)!z^{n+1}}\rho^{2n}J_{n+1}\left(z\right),
\label{eq:sr}
\end{eqnarray}
which converges for any combination of $\rho$ and $z$ (see also Bode et al. 1985).

Further more, using the expansion of $A\left(z,\rho\right)$ for the limiting case of $\rho\ll1$,
we can easily deduce the $A\left(\epsilon,z,a\right)$ for RG approximation that is generally used (Overbeck 1965; Alcock \& Hatchett 1978; Shao \& Dai 2007), which can be given by
\begin{eqnarray}
A=i\rho\left({\frac{\pi}{2z^3}}\right)^{\frac{1}{2}}J_{\frac{3}{2}}\left(z\right)=i{\frac{\rho}{z}j_1(z)},
\label{eq:rg}
\end{eqnarray}
where $j_1(x)=\sin(x)/x^2-\cos(x)/x$ is a spherical Bessel function of the first kind.
Alternatively, for large $\rho$, the approximate series expansion of the real part of the amplitude function is
\begin{eqnarray}
{\rm Re}A_{{\rm large}}=\frac{J_{1}\left(z\right)}{z}+\rho \frac{1}{y^2}
\left(\frac{\pi y}{2}\right)^{\frac{1}{2}}Y_{\frac{3}{2}}\left(y\right)+\sum_{n=1}^{\infty}\frac{z^{n-1}}{\rho^{2n}}\frac{\left(2n\right)!}{2^n n!}J_{n-1}\left(z\right),
\label{eq:lr}
\end{eqnarray}
where $Y_{\alpha}(x)$ is the Bessel function of the second kind (see the Appendix A for details). The series expansion for large $\rho$ converges when $z\ll\rho$ holds.

Above, we mentioned three approaches for calculating the amplitude function within the AD approximation. The first one is using the exact integral form defined by Equation~(\ref{eq:int}), which hereafter we label as the ``accurate AD''. The second and third ones are the approximate series expansions, which correspond to two limiting cases when calculating the real part of the amplitude function as given by Equations~(\ref{eq:sr}) and (\ref{eq:lr}), which hereafter we label as the ``series expansion for small $\rho$'' and ``series expansion for large $\rho$'', respectively. In Figure~\ref{fig2} we show the normalized differential cross sections calculated by these three approaches of the AD approximation compared with RG approximation for different $\rho$. We can obviously distinguish the difference between the different methods. There are two cases when the approximations are obviously inaccurate and the evaluated cross sections deviate dramatically from the accurate ones, so they are shown in the inset panels. As it has turned out, the approach of accurate AD is very time-consuming even though still feasible in acquiring the cross section as shown in Figure 2. But it is no longer a practical option in the following multiple integrals since it engages complex integral which will be troublesome as further used as an integrand. So we discuss the accuracy of the approximate series expansions compared with the accurate complex integral and provide a practical approach to calculate the differential cross section in the following.

{\it Series expansion for small $\rho$}: the top panel of Figure 3 shows a contour of the normalized difference at the first peak of the differential cross section between the accurate AD and the series expansion for small $\rho$. We can find that the number of terms $n$ in summation does not affect the accuracy once it is $\geq 50$. In particular, for the accuracy better than $0.01\%$ at the first peak of the differential cross section, we must have $\rho \leq30$ and $n \geq 50$ when using the series expansion for small $\rho$. Therefore in our work we adopt $n=50$ whenever the series expansion for small $\rho$ is employed.
The left panel of Figure 4 shows a contour of the normalized difference of the differential cross section at different $z$ and $\rho$ between the series expansion for small $\rho$ and accurate AD. Obviously, series expansion for small $\rho$ is valid only when $\rho<z$. By the contour in the left panel of Figure 4, the valid boundary could be well defined and approximated by $z\geq3.75\rho-75$ (the dashed line) for the accuracy better than $0.01\%$.

{\it Series expansion for large $\rho$}: the bottom panel of Figure 3 also shows a contour of the normalized difference at the first peak of the differential cross section between the accurate AD and the series expansion for large $\rho$ with different number of terms $n$ in summation. We can find that the number of terms $n$ doesn't have significant influence on the precision and in contrary to the case of small $\rho$, $n=1$ has the highest accuracy for given $\rho$. So in the following calculation, we adopt $n=1$ for the series expansion for large $\rho$ . One would also expect the series expansion for large $\rho$ is valid when $\rho>z$. The right panel of Figure 4 plots a contour of the normalized difference of the differential cross section at different $z$ and $\rho$ between the accurate AD and the series expansion for large $\rho$ with $n=1$. We can find that the accuracy is better than $0.01\%$ in the boundary $z<3.75\rho-75$.

Therefore, for convenience in computation (with an accuracy better than $0.01\%$), we can evaluate the differential cross section with series expansion for either small $\rho$ ($n=50$) or large $\rho$ ($n=1$) instead of the accurate AD. The boundary can be explicitly set at $z=3.75\rho-75$.

\section{Temporal and Spectral Properties}
\label{sec:model}

In the following we apply our approach for evaluating the differential cross section within AD approximation to investigating the X-ray scattering scenarios in GRBs (Shao \& Dai 2007; Shao et al. 2008). Accordingly, the light curve of scattered X-ray emission in a given energy band, e.g., $0.3-10$ keV for Swift X-Ray Telescope (XRT), can be given by
\begin{eqnarray}
F^{\rm echo}(t)=\int_{E_-}^{E_+}F^{\rm echo}_{E}(t){\rm d}E,
\end{eqnarray}
where the flux density, i.e., the time-dependent spectrum, $F^{\rm echo}_{E}(t)$ can be given by
\begin{eqnarray}
F^{\rm echo}_{E}(t)=\int_{a_-}^{a_+}
S(E)\frac{{\rm d}N}{{\rm d}a}\frac{c\pi a^2}{R}\left(\frac{2\pi E a}{hc}\right)^2\left|A(\hat{z},\rho)\right|^2{\rm d}a.
\end{eqnarray}
Here we define $\hat{z}(t)=(2\pi Ea/hc)\sqrt{2(1+z_{\rm RS})ct/R}$, $z_{\rm RS}$ is the redshift of the source, $R$ is the distance of dust screen to the GRB source, and $a_+$ and $a_-$ are the maximum and minimum radius of the dust grain, respectively. $S(E)$ is the initial spectrum function of the GRB source which, in the gamma-ray band, has the form of
\begin{equation}
S(E)=C\left({E\over{100\,{\rm keV}}}\right)^\delta {\rm
exp}\left[-{(\delta+1)E\over E_{\rm p}}\right],
\end{equation}
where $C$ is a constant coefficient and $\delta$ and $E_{p}$ are the spectral index and the observed peak energy of the source (Preece et al. 2000). For X-ray band, we adopt the typical values $E_{p}\simeq200\,{\rm KeV}$, but $\delta$ is not necessarily equal to $\sim0$ and could be taken as a free parameter. ${\rm d}N/{\rm d}a$ is the column density per unit size of the dust grain, and can be given by the power-law form
\begin{eqnarray}
\frac{{\rm d}N}{{\rm d}a}=Da^\beta \qquad (a_{-}\le a \le a_{+}),
\end{eqnarray}
where $D$ is a constant coefficient and $\beta$ is the power-law index (Mathis et al. 1977).
In the following, we note that there are some parameters that can influence the light curves and hardness ratios. We analyse the effect of the large grains, the distance from the dusty shell to the GRB site and the index of the grain size.

Figure~5 plots the light curves and `pseudo' spectral index derived by AD and RG approximations, respectively (see Shen et al. 2009 for the definition of the `pseudo' spectral index). For demonstration of our calculations here, an arbitrary choice of constants $C$ and $D$ has been adopted. We will come to deal with the normalization later in Section 4. We can find that, for size distribution of typical interstellar grains between $0.005\,{\rm \mu m}$ and $0.3\,{\rm \mu m}$, the light curves and spectra are undistinguishable using both approximations. However, when the maximum value of the size of grains extends over $1\,{\rm \mu m}$, the discrepancy is very obvious, especially in the early phase. We can understand the discrepancy from the different differential cross sections evaluated by these two approximations. RG approximation assumes that all the dust grains have the same normalized differential cross section regardless of the size of the dust grains. However, the normalized differential cross sections calculated by AD approximation are in fact very sensitive to the size of the dust grains. Figure 2 shows that the normalized differential cross section calculated by AD approximation is smaller than the one derived by RG approximation when $z$ is relatively small and $\rho$ is very large. So the light curves derived form AD approximation are smaller than the ones calculated by RG approximation at the early phase when the the light curve is dominated by the first peak in the differential cross section. At later time, both the significantly-softening spectra and steeply-decaying light curves would be the same for AD and RG approximations.

Further more, the detailed spectral differences are shown in Figure 6. Compared with the spectra calculated by RG approximation, the spectra calculated by AD approximation all have an increase in the lower energy when the size of the grains is larger which causes the spectra harder. This could be understood with the results shown in Figure 2. For the fixed size of dust grains, the lower the photon energy is, the larger the phase shift is, which makes the normalized differential cross section calculated by AD approximation is smaller than the one derived by RG approximation. So the lower energy part of the spectrum calculated by RG approximation would be overestimated. A similar result has been discussed in Smith \& Dwek (1998). As the lower energy part of spectrum is overestimated, the spectral index evaluated by RG approximation in early phase will be significantly larger than the one derived by AD approximation as shown in Figure~5. Therefore, unlike RG approximation, the spectral profile given by AD approximation is less-dependent on the size of dust grains. As presented in the next section, a detailed modeling for the observational data of both lightcurves and spectra will help determine the real size information of dust grains. In particular, it is suggested by both the RG and AD approximations that an early plateau in the lightcurve of X-ray afterglow would imply smaller sizes of dust grains.

The effect of distance $R$ of the dust screen from the GRB source on the light curves have been studied by Shao \& Dai (2007), which are also revised here with AD approximation and the results are similar. The distance $R$ is usually considered as approximately tens of parsecs (Garc\'{i}a-Segura et al. 1996; Mirabal et al. 2003; Dwarkadas 2007; Shao et al. 2008). If $R$ decreases, the time scale in the system shrinks, which makes the light curve and spectrum have a ``time shift'' to the early phase, so the light curve would appear brighter at the early phase and the period of significant spectral softening would appear earlier.

Besides the distance $R$, the power-law index $\beta$ of the size distribution of dust grains can also reflect the evolution of the dust grains. When the dusty screen is at a distance about tens of parsecs, the optical-UV flash can sublimate the small dust grains (Waxman \& Draine 2000). This mechanism can make $\beta$ larger or the distribution flatter. As shown in Figure~7, $\beta$ significantly changes the scaling of the light curve without affecting its shape too much. We may directly realize this by turning to Equation (10). When the maximum value of the grain size $a_+$ is smaller than $1 {\rm \mu m}$, $a^\beta$ decreases if $\beta$ increases, which makes the light curve dimmer. On the contrary, when $a_+$ is larger than $1\,{\rm \mu m}$, $a^\beta$ increases if $\beta$ increases, which makes the light curve brighter. For the spectral index, $\beta$ may have more notable influence on its evolution.

\section{Spectral Softening in GRB~130925A}
\label{sec:GRB130925A}

The major feature of this X-ray scattering scenario is the significant spectral softening as has been mentioned in the previous works (Shao \& Dai 2007; Shao et al. 2008; Shen et al. 2009). This has been manifested in GRB 090417B which fits nicely both the temporal and spectral features as predicted by the original model (Shao and Dai 2007), and the dark nature of its optical afterglow directly indicates the existence of large amount of dust grains in the vicinity of the GRB sites (Holland et al. 2010). Recently, the X-ray afterglow of GRB~130925A also showed the similar spectral softening which was not expected by the standard afterglow models.

GRB~130925A was triggered on 2013 September 25 at 04:09:27 UT by Fermi Gamma-Ray Burst Monitor (GBM; Fitzpatrick et al. 2013) and at 04:11:24 UT by the Swift Burst Alert Telescope (BAT; Lien et al. 2013). The X-ray afterglow was detected by Swift XRT at $147.4\,{\rm s}$ after the BAT trigger with no early optical counterpart in Swift UVOT (Lien et al. 2013) or in ground TAROT robotic telescope (Klotz et al. 2013) at the same location. An uncataloged optical source was found in the XRT error circle (Sudilovsky et al. 2013a) with emission lines at the redshift of $z=0.347$ (Vreeswijk et al. 2013; Sudilovsky et al. 2013b). The optical source did not show variation in brightness and suggest itself as the host galaxy of GRB~130925A (Butler et al. 2013a, 2013b; Holland 2013). The non-detection of decaying optical afterglow renders GRB~130925A a dark GRB which is another similar feature with GRB~090417B (Holland et al. 2010).

The third feature that makes GRB~130925A similar with GRB~090417B is its excessive early X-ray flares that suggests a long prompt duration (Burrows et al. 2013; Evans et al. 2013), which also trigger MAXI about an hour after the Swift BAT trigger (Suzuki et al. 2013). Burrows et al. (2013) proposed the interesting possibility that this burst could be another tidal disruption event like Swift~J1644+57/GRB~110328A (Burrows et al. 2011; Levan et al. 2011; Zauderer et al. 2011; Bloom et al. 2011). It is worth to mention that, the similarity between GRB~090417B and Swift~J1644+57/GRB~110328A in the early phase has already been noticed (Shao et al. 2011). But the late-time temporal behaviour of GRB~130925A is typical of long-duration GRBs (Malesani et al. 2013) as shown in the upper panel of Figure~8 and not likely a tidal disruption event since it is located relatively far from the center of the host galaxy (Tanvir et al. 2013). Recently, the long-lasting prompt emission is proposed to originate from hyperaccretion disk around a precessed stellar black hole (Hou et al. 2014).

The main feature of GRB~130925A that makes it different from typical long-duration GRBs is its significant late-time spectral softening as shown in the lower panel of Figure~8. For details of how these light curves and spectral analysis were produced, see Evans et al. (2009;2010). A very recent study also showed that an additional spectral component is required in late-time afterglow of GRB~130925A (Bellm et al. 2014). While this spectral feature is a several challenge to standard afterglow models (e.g., Holland et al. 2010), it has been well predicted by the X-ray scattering scenario. At this stage, we find that this scenario could well reproduce both the late-time temporal and spectral features of GRB~130925A\footnote{Amazingly, an independent paper (Evans et al. 2014) that is also devoted to this idea appeared on arXiv.org at the same time with our manuscript.}. Given the early plateau showing in the x-ray light curve after several $10^4$ s, a relatively smaller size of dust grains is suggested according to the results shown in Figure 5. In this case, AD and RG approximations would more or less agree with each other.

As have been discussed in last section, several physical parameters of the dust grains would affect the resulting light curves and spectra. To show how well this X-ray scattering scenario would reproduce the X-ray afterglow of GRB~130925A and provide a first-hand quantitative study of the model parameters that have played a major role in determining both the light curves and spectra, we adopt fixed values of $a_{-}=0.005\,{\rm \mu m}$, $\beta=-3.5$ and $\delta=-0.3$ and only let $a_{+}$ and $R$ vary independently. Then we try to fit the temporal and spectral data simultaneously and proceed with AD and RG approximations, respectively. The best-fit model parameters for AD approximation (solid lines in Figure~8) are $a_{+}=0.40\pm0.10\,{\rm \mu m}$ and $R=650\pm350\,{\rm pc}$ and the fitting yields a joint reduced $\chi^2$ of 1.08 with the flux in logarithmic scale and the photon index in linear scale. Here the errors for the best-fit parameters are simply determined by locating the nearest reduced $\chi^2$'s of 1.5 in the parameter space. We assume the model is considered unfit if the corresponding reduced $\chi^2$ is larger than 1.5. Accordingly, the best-fit model parameters for RG approximation (dashed lines in Figure~8) are $a_{+}=0.30\pm0.15\,{\rm \mu m}$ and $R=600\pm300\,{\rm pc}$ and the fitting yields a joint reduced $\chi^2$ of 1.38. Therefore, according to this X-ray scattering scenarios, a typical distribution of small grains as in the interstellar medium is suggested for GRB~130925A. Though, the seemingly better fitting with a relatively smaller reduced $\chi^2$ by the AD approximation should not be overemphasized, since both approximations would be close and accurate given the inferred smaller grain size for this burst.

We can also estimate the flux ratio between the scattered X-ray emission and the prompt (the source) X-ray emission to determine the amount of dust grains that have been evolved in the scattering processes by adopting the method introduced in Shen et al. (2009). We integrate the best-fitting (flux density at 1 keV) light curve to $10^7$ sec to obtain a specific fluence of $2.3\times 10^{-6}\,{\rm erg}\,{\rm s}^{-1}\,{\rm cm}^{-2}\,{\rm keV}^{-1}$ at 1 keV for the scattered X-ray emission. According to the observed gamma-ray fluence of $5\times10^{-4}\,{\rm erg\,cm^{-2}}$ in the 20-10000 keV energy range from Konus-Wind (S.Golenetskii et al. 2013), we can estimate a specific fluence of $8.8\times 10^{-6}\,{\rm erg}\,{\rm s}^{-1}\,{\rm cm}^{-2}\,{\rm keV}^{-1}$ at 1 keV for the source X-ray emission. Given the redshift of $z=0.347$ (Vreeswijk et al. 2013; Sudilovsky et al. 2013b), we can estimate the scattering optical depth to be $\sim0.46$ at 1 keV. According to empirical relations between the X-ray scattering and optical extinction, this would imply $A_V\simeq3.1-7.7$. This suggested high optical extinction is consistent with the optically-dark nature of GRB~130925A.

It is worth to mention that, even though this dust scattering model can well reproduce both the temporal and spectral features of the X-ray afterglow of GRB~130925A especially at late time, it has been simplified by adopting an assumption that the duration of the source/prompt X-ray emission is negligible compared to that of the afterglow emission of interest. Given the long-lasting activity of the prompt emission of GRB~130925A, further investigation to consider the complex and variable temporal profile of the prompt emission are in need. In the pioneer work, Shao \& Dai (2007) simplified the long-lasting prompt emission as a beam that has a constant flux within a given time period. Based on only one-dimensional integration for a single X-ray energy with a single dust size, they implied that the temporal profile of the scattered emission would not be affected except that the overall flux level would be increased if the prompt emission is a constant beam instead of a pulse.

Here we further investigate this problem with three-dimensional integrations taking into account of both the spectral distribution of photon energy and the size distribution of dust grains. The resulting temporal and spectral profiles of the scattered X-ray emission are shown in Figure~9. The source/prompt X-ray emission has been adopted as a constant beam with a invariant spectrum given as Equation (9), which has a duration of $10^2$, $10^3$ and $10^4$ sec, respectively. For a given prompt duration, the scattered emission is only plotted after the prompt duration when the scattered emission is supposed to emerge. For the light curves, we obtain a similar result with Shao \& Dai (2007) that the accumulative scattering of long-lasting prompt X-ray emission does not change the overall temporal profile except for a little difference at the emerging of the scattered emission. For the spectral profile, the result is also similar and the difference is more clearly shown at the beginning of the emerging of the scattered emission. The results from Figure~9 reveal that the cumulative scattering of long-lasting instead of a pulsed prompt emission would affect the temporal or spectral profile only within a time period that is comparable with the prompt duration. Therefore, the effect due to the long lasting prompt emission would be negligible very soon after the emerging of the scattered emission in a logarithmic scale.  In the future, we would take into account the complex temporal and spectral profiles of the last-lasting prompt emission to account for the very early behavior of the scattering scenario with more elaborate calculations.

The location effect of dust grains from the GRB progenitor has been further investigated in Shen et al. (2009) with extended dust zone instead of a thin layer. For a power law profile of density distribution, they found that only the most dense inner part of the dust zone would play an important role over the scattering process. Even though the realistic distribution of dust grains around GRBs has remained undetermined, the thin layer assumption has been adopted to be consistent with the wind bubble structure around evolved massive stars (e.g., Garc\'{i}a-Segura et al. 1996; Mirabal et al. 2003; Dwarkadas 2007; Shao et al. 2008). Given the typical time scale of the spectral softening, the scattering is supposed to come into force at a distance of about tens of parsecs. Therefore the standard afterglow emission from external shocks more close to the progenitor would also make significant contribution to the overall behavior of the X-ray emission. This has not been considered in this work, since the current scattering scenario could already reproduce well the major temporal and spectral features for GRB~130925A. The very early gamma-ray and X-ray emission for GRB~130925A is indeed mysterious, which we believe, might be the demonstration of the external shocks. With a combination of both external shock and scattering scenarios, much more complex X-ray afterglow would be reproduced. We will address this issue systematically with more related observations including tens of GRBs in a forthcoming paper (Wang et al. 2014, in preparation).

\section{Conclusion}
\label{sec:conclusion}

In this paper we revisited the X-ray scattering around GRBs adopting AD approximation for the differential cross section. In most of the practical cases, the generally used RG approximation is no longer valid with soft X-ray photon or large grain size according to the Mie theory. Given that the original Mie theory is complex and time-consuming in computation, we provided a valid and more practical approach within the AD approximation of Mie theory to evaluate the differential cross sections. We found that for the vicinity of GRBs where larger grains might prevail, RG approximation would overestimate the flux of dust echo emission in the early light curves and also overestimate the lower part of the spectrum. Instead, by AD approximation, a steep rise in the lower part of the spectrum would be obtained especially when the maximum size of the dust grains, the most important parameter in size distribution, is significantly larger than $1\,{\rm \mu m}$. At later time, both the significantly-softening spectrum and steeply-decaying light curve would be the same for AD and RG approximations. We also found that a smaller maximum size of dust grains would be implied by both approximations if an early plateau clearly appears in the X-ray light curve. A detailed modeling for the observational data of both lightcurve and spectrum would help determine the size distribution of dust grains.

As has been previously delivered (Shao \& Dai 2007; Shao et al. 2008; Shen et al. 2009), the significant spectral softening, as a unique outcome of this X-ray scattering scenario, is generally expected with either RG or AD approximation adopted. Even though most X-ray afterglows did not exhibit such a spectral feature, it would be a great challenge to standard afterglow models if this feature indeed showed up (Holland et al. 2010). We gave an example of how this feature may have also been observed in a recent burst GRB~130925A. The significant spectral softening and optically dark nature of GRB~130925A strongly favor the X-ray scattering scenario. Given the early plateau present in the X-ray light curve, we found that a typical distribution of smaller grains as in the interstellar medium would be suggested. Therefore, the X-ray scattering scenarios with either AD or RG approximation adopted could both well reproduce the late-time temporal and spectral profile simultaneously for GRB~130925A.

\acknowledgments

This work made use of data supplied by the UK Swift Science Data Centre at the University of Leicester. We thank the referee for considerate comments which have helped to improve the manuscript. We also thank Yi-Zhong Fan for helpful discussions. This work was supported in part by the National Basic Research Program of China (No. 2014CB845800) and the National Natural Science Foundation of China (grants 11361140349 and 11103083).

\appendix

\section{APPENDIX A - BESSEL FUNCTIONS}

Bessel functions of the first kind $J_{\alpha}(x)$, for real $\alpha$, can be given by
\begin{eqnarray}
J_{\alpha}(x)=\sum_{m=0}^{\infty}\frac{(-1)^m}{m!\Gamma(m+\alpha+1)}\left(\frac{x}{2}\right)^{2m+\alpha},
\end{eqnarray}
where $\Gamma(x)$ is the gamma function and can be given by
\begin{eqnarray}
\Gamma(x)=\int_{0}^{\infty}t^{x-1}e^{-t}{\rm d}t.
\end{eqnarray}
The special case of $\alpha=0$ gives
\begin{eqnarray}
J_0(x)=\sum_{m=0}^{\infty}\frac{(-1)^m}{(m!)^2}\left(\frac{x}{2}\right)^{2m}.
\end{eqnarray}
Bessel functions of the second kind $Y_{\alpha}(x)$, for integer $\alpha$, can be given by
\begin{eqnarray}
Y_{\alpha}(x)=\frac{\cos(\alpha\pi)J_{\alpha}(x)-(-1)^{\alpha}J_{\alpha}(x)}{\sin(\alpha\pi)}
\end{eqnarray}

There are some useful relationships for integer $\alpha$:
\begin{eqnarray}
\sqrt{\frac{\pi}{2x}}J_{\alpha+\frac{1}{2}}(x)=j_{\alpha}(x),
\end{eqnarray}
\begin{eqnarray}
\sqrt{\frac{\pi}{2x}}Y_{\alpha+\frac{1}{2}}(x)=y_{\alpha}(x),
\end{eqnarray}
where $j_{\alpha}(x)$ and $y_{\alpha}(x)$ are called the spherical Bessel functions of the first and second kinds, respectively, and they can be given by
\begin{eqnarray}
j_{\alpha}(x)=(-x)^{\alpha}\left(\frac{1}{x}\frac{{\rm d}}{{\rm d}x}\right)^{\alpha}\frac{\sin(x)}{x},
\end{eqnarray}
and
\begin{eqnarray}
y_{\alpha}(x)=-(-x)^{\alpha}\left(\frac{1}{x}\frac{{\rm d}}{{\rm d}x}\right)^{\alpha}\frac{\cos(x)}{x}.
\end{eqnarray}
The mathematical functions listed here can be found in the online NIST Handbook of Mathematical Functions (http://dlmf.nist.gov/).

\section{APPENDIX B - THE PHASE SHIFT}

The phase shift can be derived using the ``Drude approximation'' (Bohren and Huffman 1983; Smith and Dwek 1998). The complex refractive index of the dust grain can be given by
\begin{eqnarray}
m=\sqrt{1-\frac{\omega_{{\rm p}}^2}{\omega^2+i\gamma\omega}},
\end{eqnarray}
where $\omega$ is the circular frequency of the photon, $\omega_{{\rm p}}^2=4\pi c^2 {\rm n}_e {\rm r}_e$ is the plasma frequency of the medium (where ${\rm n}_e$ is electron number density and ${\rm r}_e$ is the classical radius of the electron) and $\gamma$ is the damping term. Considering $|m-1|\ll1$,
the Equation (B1) can be rewritten by
\begin{eqnarray}
|m-1|\approx\frac{{\rm n}_e {\rm r}_e\lambda^2}{2\pi}\left(\frac{1}{1+i\gamma/\omega}\right),
\end{eqnarray}
where $\lambda$ is the wavelength of the photon.
Here we ignore the damping and the phase shift can be approximated as
\begin{eqnarray}
\rho=2a{\rm n}_e {\rm r}_e\lambda.
\end{eqnarray}
If we assume ${\rm n}_e=\rho_{\rm m}/2m_{\rm H}$ where $\rho_{\rm m}$ is the dust mass density, the phase shift can be given by
\begin{eqnarray}
\rho\approx\ 6 \left(\frac{\rho_{\rm m}}{3\,{\rm g\,cm^{-3}}}\right)\left(\frac{E}{1\,{\rm keV}}\right)^{-1}\left(\frac{a}{1\,{\rm \mu m}}\right).
\end{eqnarray}
If the dust layer (and the GRB) is located at a redshift $z_{\rm RS}$, the effective phase shift can be given by
\begin{eqnarray}
\rho\approx  3 \left(\frac{1+z_{\rm RS}}{2}\right)^{-1} \left(\frac{\rho_{\rm m}}{3\,{\rm g\,cm^{-3}}}\right)\left(\frac{E}{1\,{\rm keV}}\right)^{-1}\left(\frac{a}{1\,{\rm \mu m}}\right).
\end{eqnarray}

\begin{figure}
\epsscale{.80} \plotone{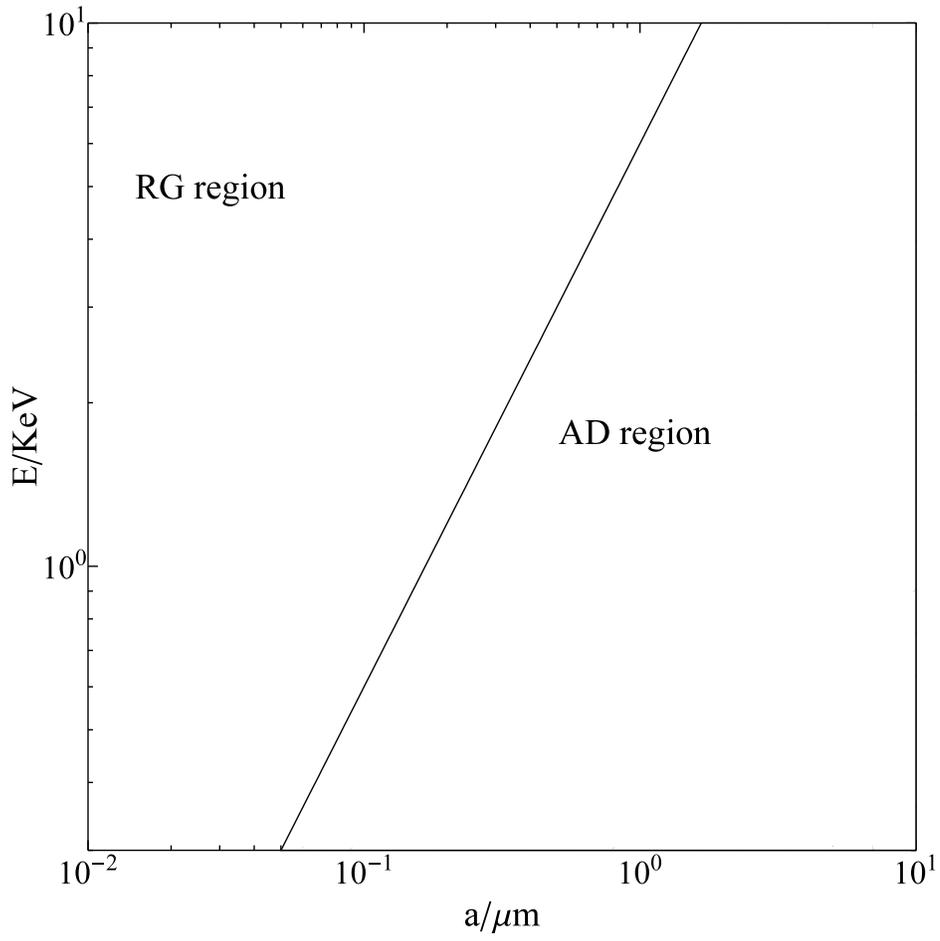} \caption{Valid regions of the photon energy and grain size for RG and AD approximations, respectively. \label{fig1}}
\end{figure}

\begin{figure}
\epsscale{.70} \plotone{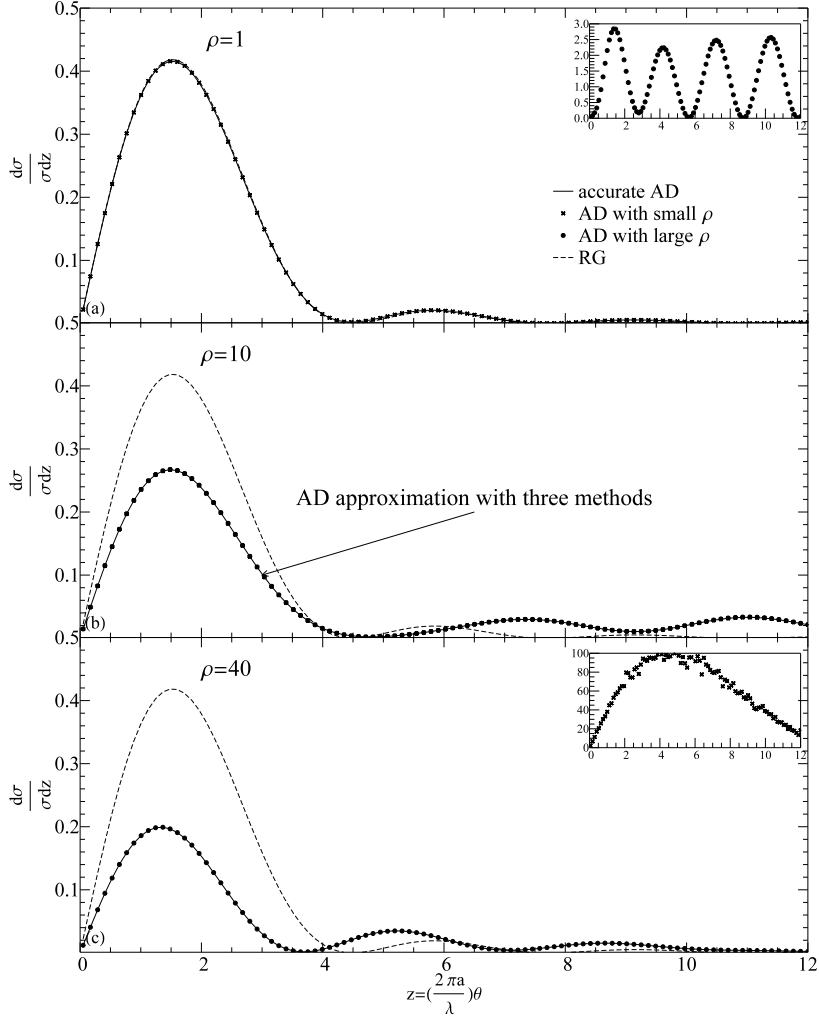} \caption{Normalized differential cross sections calculated by different approaches for different $\rho$. The dashed line shows the RG approximation as in Equation (5), the solid line shows the accurate AD approximation as in Equation (2), the filled circle shows the series expansion for large $\rho$, and the cross shows the series expansion for small $\rho$. Top: for $\rho=1$, the inset shows the normalized differential cross section calculated by the series expansion for large $\rho$; Middle: for $\rho=10$; Bottom: for $\rho=40$, the inset shows the normalized differential cross section calculated by the series expansion for small $\rho$.  \label{fig2}}
\end{figure}

\begin{figure}
\epsscale{.70} \plotone{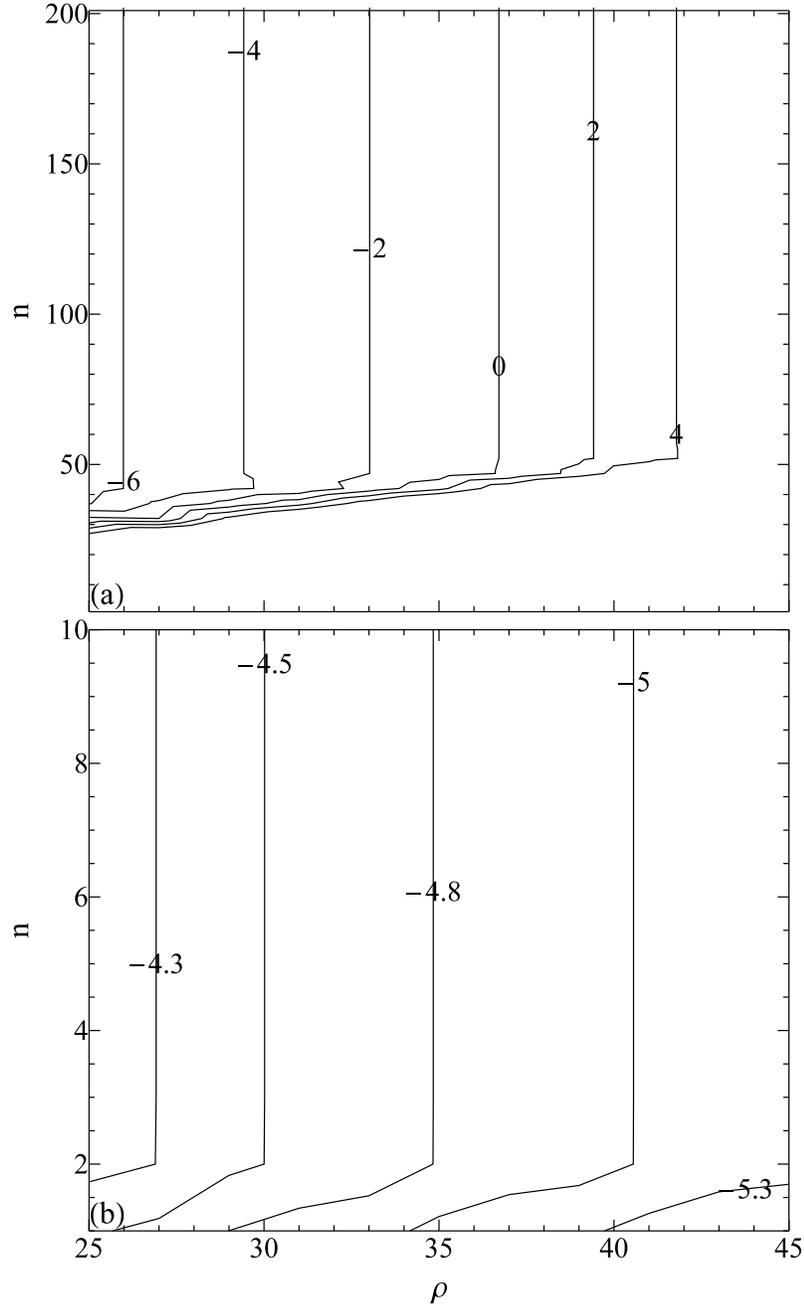} \caption{The contours (in logarithmic scale) of the normalized error at the first peak of the differential cross section compared with the accurate AD for the series expansion for small $\rho$ (Top) and large $\rho$ (Bottom) varying with different $\rho$ and $n$. \label{fig3}}
\end{figure}

\begin{figure}
\epsscale{.80} \plotone{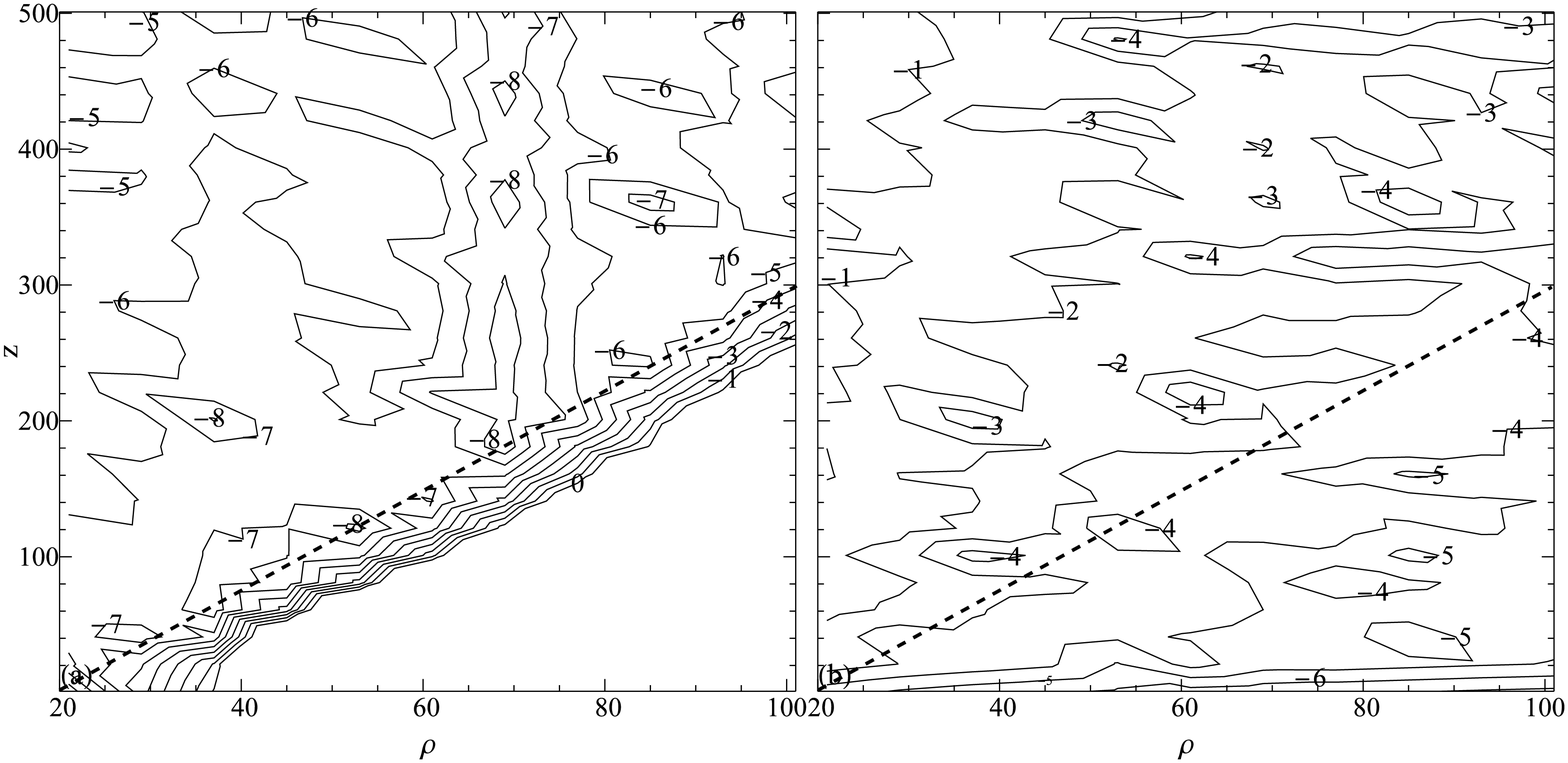} \caption{The contours (in logarithmic scale) of the normalized error in the differential cross section at different $z$ and $\rho$ between the series expansion approaches (Left: for small $\rho$; Right: for large $\rho$) and the accurate AD. The dashed line indicates $z=3.75\rho-75$ (see the text). \label{fig4}}
\end{figure}

\begin{figure}
\epsscale{.90} \plotone{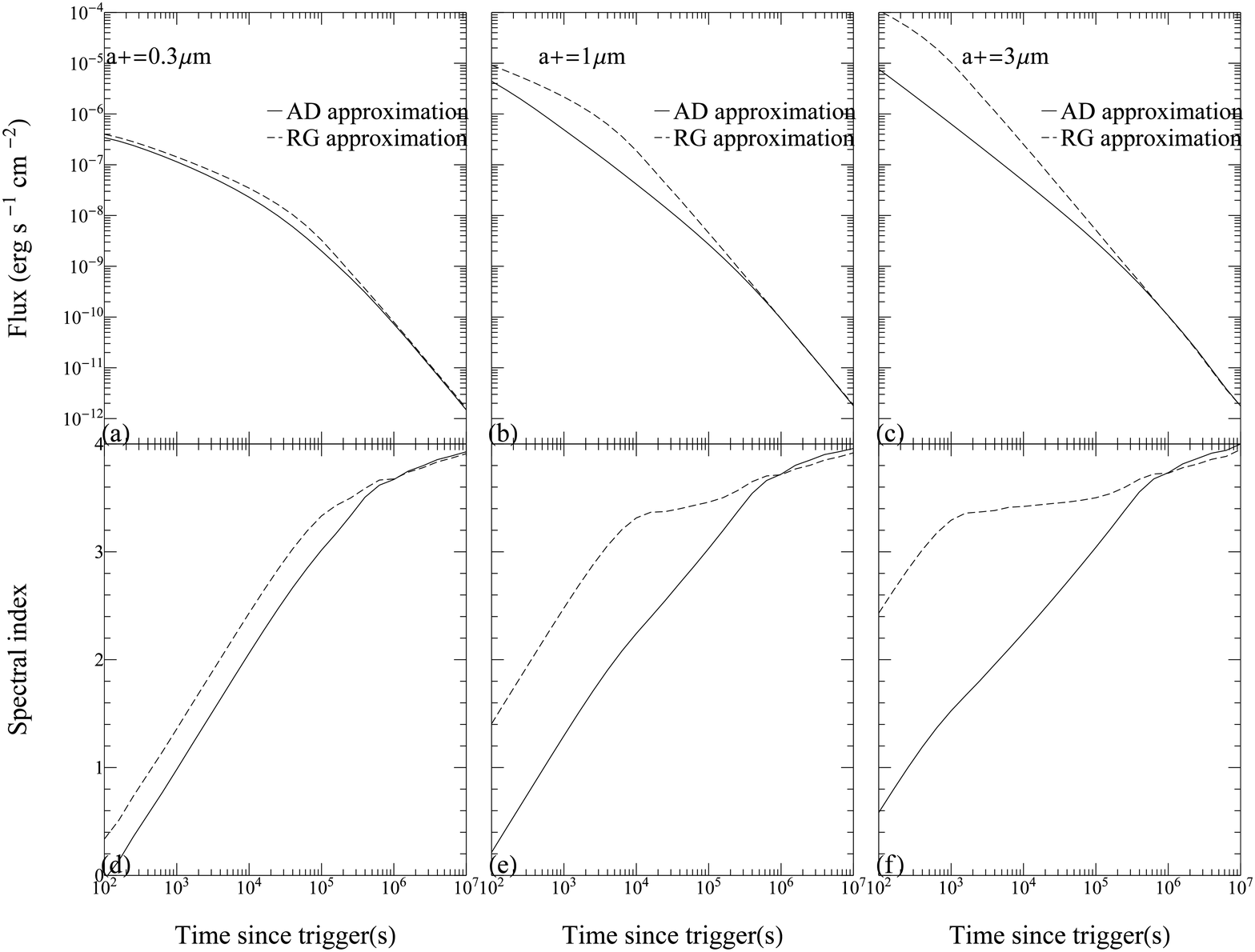} \caption{The light curves (upper panels) and the evolution of the spectral indices (lower panels) of the scattered X-ray emission by adopting the AD (solid lines) and the RG (dashed lines) approximations. From left to right, the parameter $a_{+}$ has a value of 0.3, 1, and 3$\,{\rm \mu m}$, respectively. Other parameters are $R=100\,{\rm pc}$, $\beta=-3.5$, $\delta=0$, $z_{\rm RS}=1$ and $a_{-}=0.005\,{\rm \mu m}$. \label{fig5}}
\end{figure}

\begin{figure}
\epsscale{.75} \plotone{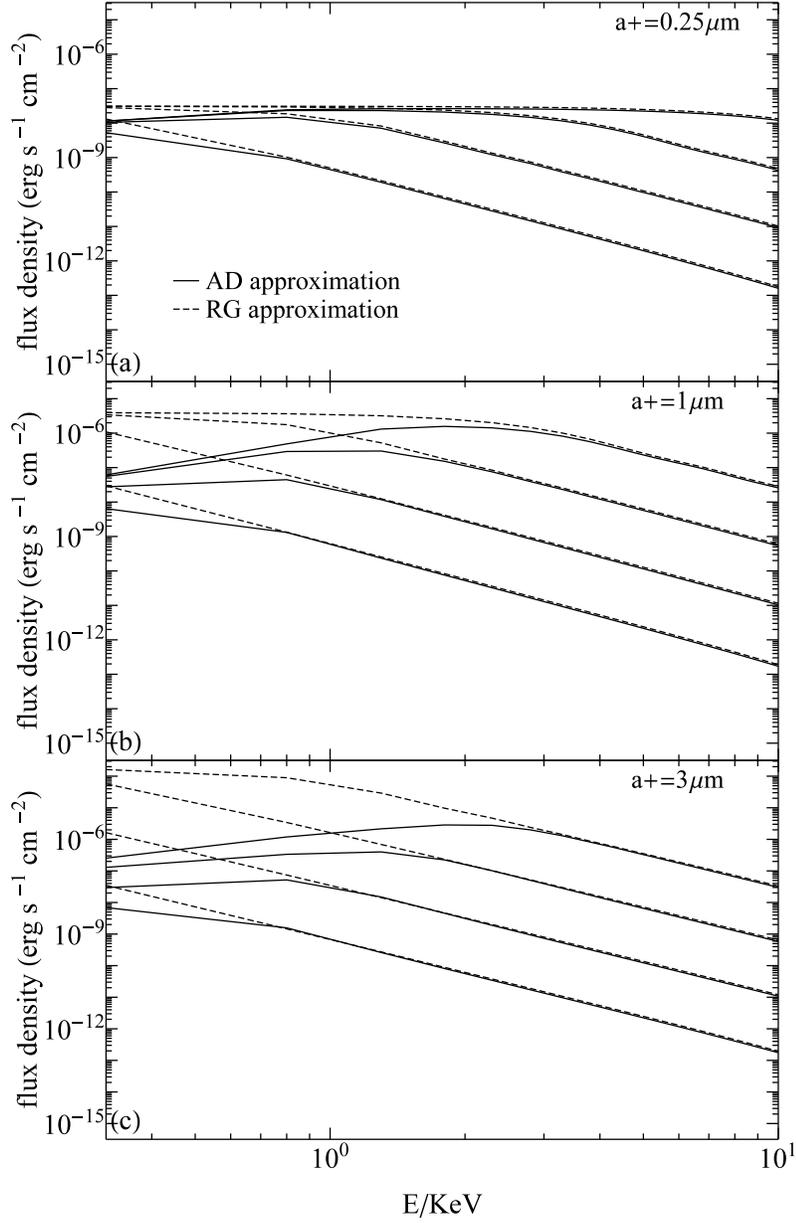} \caption{The spectral of X-ray echo emission derived from both the AD and RG approximations. The spectral are evaluated at four times since the burst, which are $10^2\,{\rm s}$, $10^3\,{\rm s}$, $10^4\,{\rm s}$, $10^5\,{\rm s}$ respectively from the top to the bottom. Other parameters: $\beta=-3.5$, $\delta=0$, $R=100\,{\rm pc}$, $z_{\rm RS}=1$, $a_{-}=0.005\,{\rm \mu m}$ and $C=10^{-4}\,{\rm erg}\,{\rm s^{-1}}\,{\rm cm^{-2}}$. \label{fig6}}
\end{figure}

\begin{figure}
\epsscale{.80} \plotone{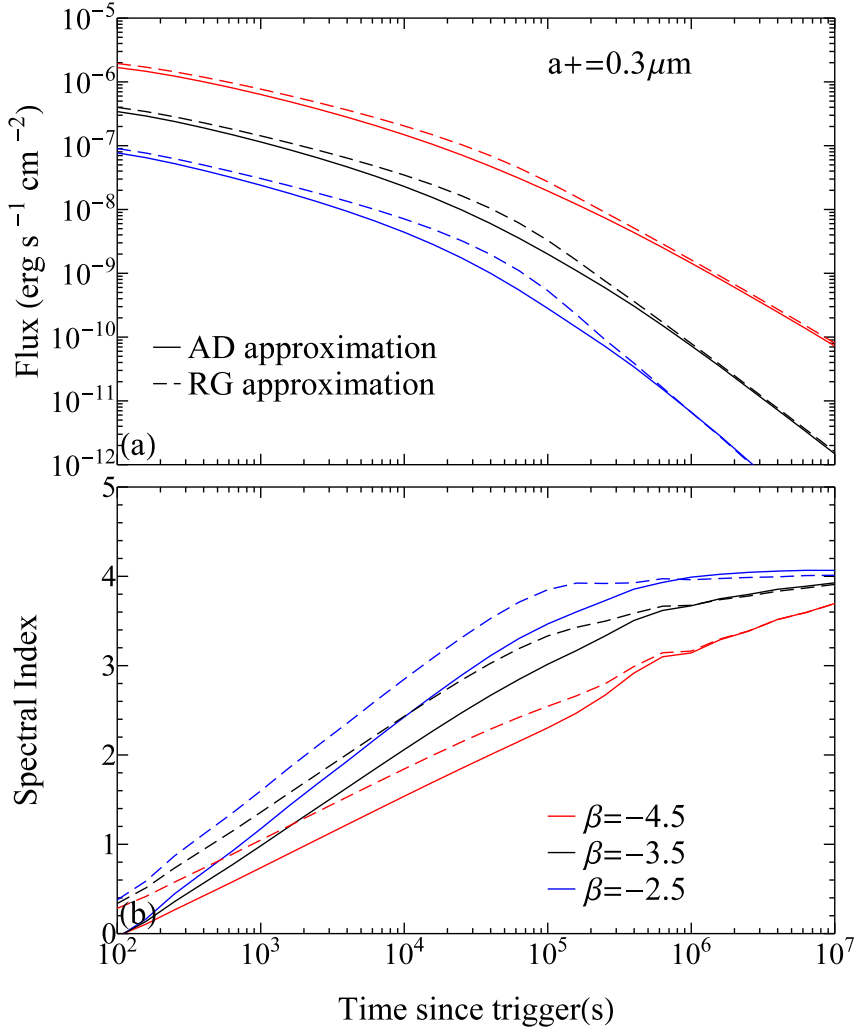} \caption{The light curves (upper panel) and the evolution of the spectral indices (lower panel) of the scattered X-ray emission with different size index $\beta$. The results adopting the AD and the RG approximations are indicated with solid lines and dashed lines, respectively. In each panel, the blue, black and red lines represent different size indices of $\beta=-2.5$, $\beta=-3.5$ and $\beta=-4.5$, respectively. Other parameters adopt the values of  $a_{-}=0.005\,{\rm \mu m}$, $a_{+}=0.3\,{\rm \mu m}$, $R=100\,{\rm pc}$, $\delta=0$ and $z_{\rm RS}=1$.\label{fig7}} \end{figure}

\begin{figure}
\epsscale{.80} \plotone{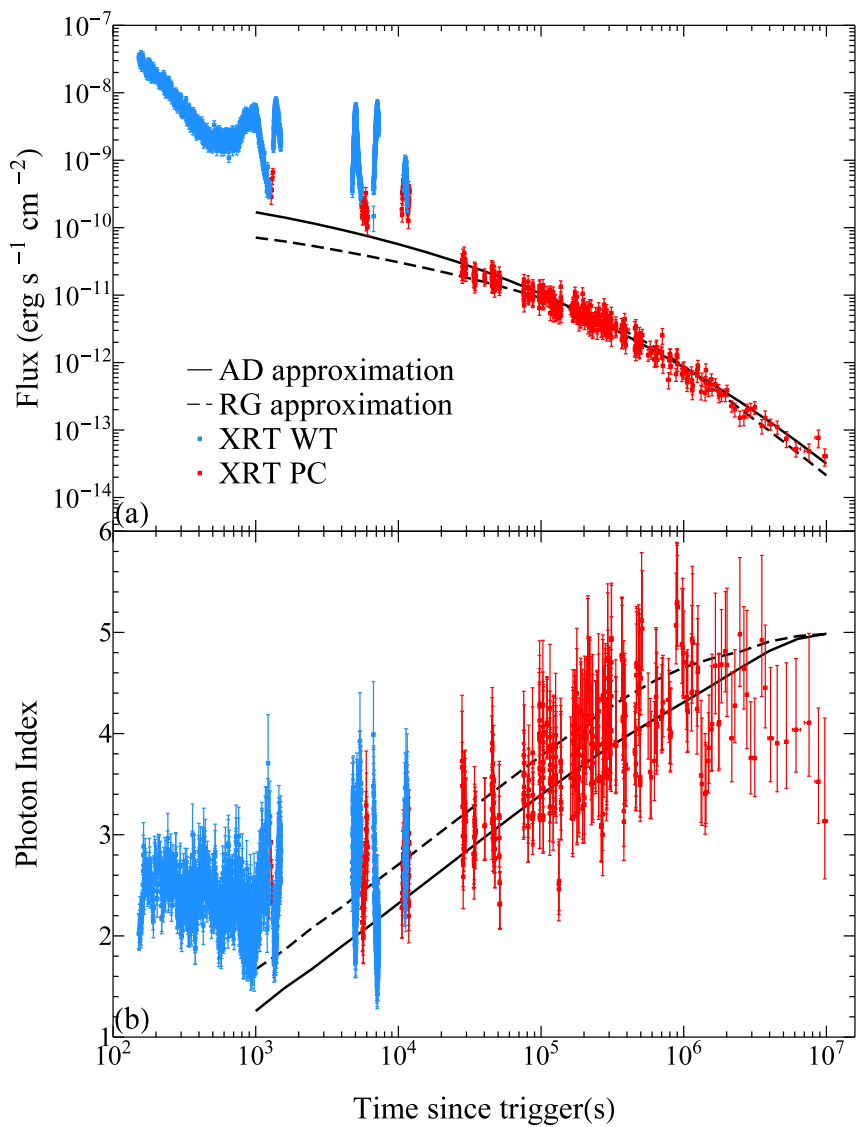} \caption{The light curve (upper panel) and the evolution of the spectral indices (lower panel) of the X-ray afterglow of GRB~130925A detected by Swift XRT. The data from WT and PC modes of XRT are plotted in blue and red squares, respectively. The solid and dashed lines indicate best-fit models adopting AD (solid lines) and RG (dashed lines) approximations, respectively. \label{fig8}}
\end{figure}

\begin{figure}
\epsscale{.80} \plotone{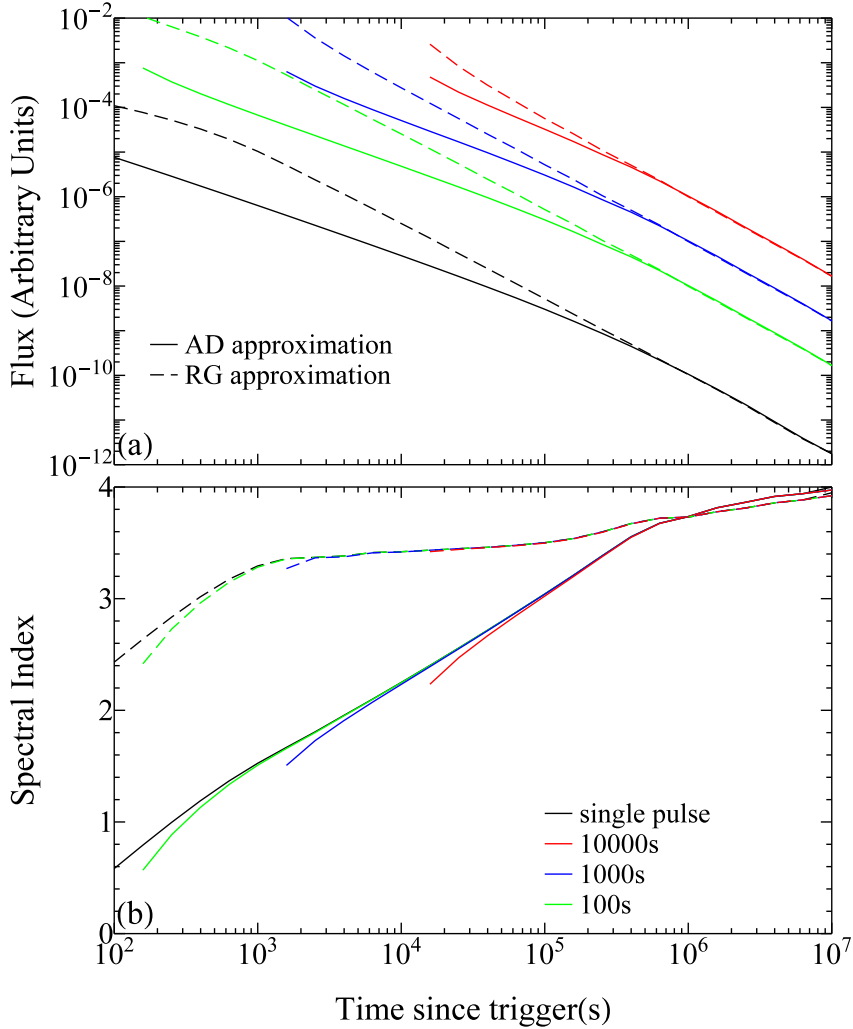} \caption{The light curves (upper panel) and the evolution of the spectral indices (lower panel) of the scattered X-ray emission by taking into account of different durations of the source/prompt emission. The results adopting the AD and the RG approximations are indicated with solid line and dashed line, respectively. Black line represents the limiting case of pulsed prompt emission. From the top to the bottom in each panel, the red, blue and green lines represent different source/prompt durations which are $10^4$, $10^3$ and $10^2$ sec, respectively. Other parameters adopt the values of $R=100\,{\rm pc}$, $\beta=-3.5$, $\delta=0$, $z_{\rm RS}=1$, $a_{-}=0.005\,{\rm \mu m}$ and $a_{+}=3\,{\rm \mu m}$.\label{fig9}}
\end{figure}

\end{document}